\documentclass[11pt]{article}
\usepackage{fleqn}
\usepackage[dvips]{graphicx}
\begin{document}

\title
{\bf 
Interference pattern of the Coulomb and the strong Van der Waals forces 
 in p-p scattering\footnote{NUP-A-2000-13}
}
\vspace{20mm}
\author{Tetsuo Sawada \\
{\small \em Atomic Energy Research Institute, Nihon University, 
Tokyo, Japan 1010062}
\thanks{Associate member of AERI for research. \ \ \  
  e-mail address: t-sawada@fureai.or.jp}}
\date{}

\maketitle
\thispagestyle{empty}
\vspace{95mm}
\begin{flushleft}
{\large\bf Abstract}
\end{flushleft}
    In order to confirm the strong Van der Waals force in the 
nucleon-nucleon interaction, it is proposed to measure precisely 
the angular  distribution of the cross section of the low energy  
($T_{lab}=20 \sim 30$MeV.) proton-proton scattering.   By using 
the spectrum of the long range interaction obtained from the 
analysis of the phase shift data of the S-wave of the p-p scattering,   
a characteristic interference pattern, which arises from the 
repulsive Coulomb and the attractive strong Van der Waals forces, 
is predicted.  The pattern has a dip at $\theta_{c.m.}= 14^{\circ}$ 
with the depth around one per cent.

\newpage
\setcounter{page}{1}

\section{Introduction }

\ \ \ \ \  In the composite model of hadron,\cite{dyon}
 in which the fundamental 
constructive force is a  strong or super-strong Coulombic force, 
the quantum fluctuation of the composite states causes the strong 
Van der Waals interaction between hadrons.   Historically  before 
1960's, hadrons were regarded as elementary particles and the 
interactions arose from the exchange of mesons with masses, therefore   
the forces were inevitably short range.   After the introduction of 
the composite model of hadron, the idea of the short range force 
is taken over to the new hadron physics, because when the momentum 
transfer is small and we do not explore the inside of hadrons, there 
must be no differences whether the hadrons are composite or 
elementary.   Although the short range nature of the interactions 
between hadrons are widely believed, it cannot be true when the 
strong Van der Waals forces are acting between hadrons.\cite{matom}
  The purpose 
of the present paper is to propose an experiment to confirm the 
existence of the strong Van der Waals interaction between nucleons 
by observing the characteristic interference in the low energy 
(20 $\sim$ 30 MeV.) proton-proton scattering.\cite{kiang} 
  The interference 
pattern arises from the destructive interference between the repulsive 
Coulomb and the attractive strong Van der Waals forces, and the cross 
section $\Delta \sigma / \sigma$ has a narrow dip in the neighborhood 
of $\theta_{c.m.}= 14^{\circ}$ with the depth around one per cent.

       In the confirmation of the strong Van der Waals interaction, 
it is desirable to have recourse to the difference of analytic 
structure of the scattering amplitude $A(s,t)$ in the neighborhood 
of $t=0$.   For the case of the short range force $A(s,t)$ 
is regular at $t=0$ and the nearest singularity occurs at $t=t_{min}$ 
, where $(t_{min})^{1/2}$ is the smallest mass exchanged in the 
t-channel.   On the other hand when the long range force is acting 
and the asymptotic form of the potential is $V(r) \sim -C/r^{\alpha}$ 
, an extra singularity occurs at $t=0$ in $A(s,t)$  and whose spectral 
function behaves as $A_{t}(s,t)= C' t^{\gamma} +\cdots$ in the 
threshold region.   In the next section, we shall see that the 
powers $\alpha$ and $\gamma$ are related by $\alpha=2 \gamma +3$.   
Therefore to observe the extra singularity at $t=0$ implies the 
confirmation of the long range force.   Moreover since $t=0$ is the 
end point of the physical region $-4 \nu \leq t \leq 0$, we can 
recognize the extra singularity, when the sufficiently precise data 
are available, without making any analytic continuation. 

     In particular if the long range interaction is the Van der Waals 
potential of the London type ($\alpha=6$), the amplitude $A(s,t)$ 
must have a singular term $ C' (-t)^{3/2}$.   Since $-t=2 \nu (1-z)$,  
there are two ways to observe the extra singularity of the amplitude 
$ (2 \nu)^{3/2} (1-z)^{3/2}$.  The first one is to make the partial 
wave projection and to observe the singular threshold behavior 
$\nu^{3/2}$ in the partial wave amplitudes $a_{\ell}(\nu)$.   
The second one is to fix $\nu$ and to observe the 
anomalous angular dependency $(1-z)^{3/2}$, which has a singularity 
at $z=1$.    By analysing the once subtracted S-wave amplitude of the 
p-p scattering $(a_{0}(\nu)-a_{0}(0) )/\nu$, the long range force 
was searched in the previous paper,\cite{prev} and we observed a cusp 
of the form 
$( c_{0} -c_{1} \sqrt{\nu})$ at $\nu=0$ which was characteristic to 
the Van der Waals interaction of the London type.   In section 2, we 
shall briefly review the previous search of the strong Van der Waals 
force using the once subtracted partial wave amplitude, and the 
parameters of the long range interaction will be given.   
In section 3, by using the parameters of the long range force, 
the anomalous 
angular distritution of the amplitude of the p-p scattering 
is computed, and the 
characteristic interference pattern of the cross section is 
predicted.   Section 4 will be used for remarks and comments.

\section{ Search for the strong Van der Waals force in the 
 S-wave amplitude }
\ \ \ \ \  When the scattering amplitude $A(s,t)$ has the extra 
singularity $C'(-t)^{\gamma}$, the partial wave amplitudes 
$a_{\ell}(\nu)$ also have extra singularities at $\nu=0$ and the 
threshold behaviors become $ C''_{\ell} \nu^{\gamma}$, where 
$ C''_{\ell} $ are proportional to $C'$.     Since in the hadron 
physics the data of the S-wave phase shift of the p-p scattering 
are prominent in their 
accuracy and the scattering length is determined very precisely, 
we shall analyse the once subtracted S-wave amplitude $ (a_{0}(\nu)-
a_{0}(0) )/\nu$ in search for the extra singularity $C''_{0} 
\nu^{\gamma -1}$. 
Our normalization of the partial wave amplitude $a_{0}(\nu)$ is
\begin{equation} 
   a_{0}(\nu)=\frac{\sqrt{m^2+\nu}}{X_{0}(\nu)- i \sqrt{\nu}}
\end{equation}
for the scattering of neutral particles, and $X_{0}(\nu)$ is the 
effective range function $\sqrt{\nu} \cot \delta_{0}(\nu)$ 
which is regular at $\nu=0$ and accepts 
the Taylor expansion of $\nu$.
  In order to observe the singular behavior at 
$\nu=0$, we must remove the known nearby singularities.  First of all 
we have to remove the unitarity cut by constructing a function 
\begin{equation} 
\frac{K_{0}(\nu)}{\nu} \equiv \frac{a_{0}(\nu)-a_{0}(0) )}{\nu} -
\frac{1}{\pi} \int_{0}^{\infty} \frac{ {\rm Im } a_{0}(\nu')}
{\nu' (\nu'-\nu)} d \nu'
\end{equation} 
, which is free from the right hand cut and is called the Kantor 
amplitude.\cite{kantor}  
 To facilitate the search of the extra singularity it is 
desirable to remove also the cut of the one-pion exchange (OPE).  The 
procedure is the same as the case of the unitarity cut, namely 
first compute an integration 
\begin{equation}   
\frac{K_{0}^{1 \pi}(\nu)}{\nu} \equiv \frac{1}{\pi} 
\int^{-1/4}_{-\infty} \frac{ {\rm Im } a_{0}^{1 \pi}(\nu')}
{\nu' (\nu'-\nu)} d \nu'=\frac{1}{4} \frac{g^2}{4 \pi} \{ \frac{1}
{4 \nu} \log (1+4 \nu) -1 \}
\end{equation} 
, and then subtract it from the amplitude.  In the calculation, 
the neutral pion mass is set equal to 1, and throughout this paper 
we shall use the  neutral pion mass as the unit of the energy and 
the momentum.   Since the two-pion exchange spectrum starts slowly 
at $\nu=-1$, the function $(K_{0}(\nu)-K_{0}^{1 \pi}(\nu))/\nu$ must 
be almost constant and have small slope in the neighborhood of $\nu=0$ 
when the long range interactions are absent.    On the other hand, 
for the Van der Waals interaction of the London type ( $\alpha=6$ ) 
$\gamma=3/2$ and $(K_{0}(\nu)-K_{0}^{1 \pi}(\nu))/\nu$ must 
have a singular term $C''\nu^{1/2}$, therefore we can observe a cusp 
at $\nu=0$ as long as the coefficient $C''$ is not very small. 
In this way we can examine the long range force if the $\pi$-N coupling 
constant $g^2/4 \pi$ and the S-wave phase shifts are given.  
Because the effects of the Coulomb and the vacuum polarization 
potentials are not considered, this method is applicable to the 
case where the Coulomb interaction is not important such as to the 
neutron-neutron scattering.

     Let us turn to the proton-proton scattering, where the 
vacuum polarization as well as the Coulombic interactions are 
important.  The Kantor amplitude introduced in Eq.(2) still has 
the left hand cuts, namely the 
Coulombic cut in $ -\infty < \nu \leq 0 $ and the cut of the vaccum 
polarization in $-\infty < \nu \leq -m_{e}^2$.    The difficulties are 
by-passed if we use the modified effective range function 
$X_{0}(\nu)$ of the proton-proton scattering, which is regular at 
$\nu=0$ and accepts the effective range expansion, when all the 
forces are short range except for the terms of the Coulomb and of 
the vacuum polarization.   The modified effective range function 
$X_{0}(\nu)$ for the phase shift $\delta_{0}^{E}(\nu)$ is
\begin{equation}  
 X_{0}(\nu)= \frac{C_{0}^2 \sqrt{\nu}}{1-\phi_{0}} \{ (1+\chi_{0}) 
 \cot \delta_{0}^{E} -\tan \tau_{0} \} +m e^2 h(\eta) +m e^2 
 \ell_{0}(\eta) \quad .
\end{equation}
 In Eq.(3), two well-known functions with the Coulombic order of 
 magnitudes appear, they are expressed using a new variable 
 $\eta=m e^2/(2 \sqrt{\nu})$ :
\begin{equation}   
C_{0}^2 = \frac{2 \pi \eta}{ e^{2 \pi \eta}-1} \qquad \; and \qquad 
\; h(\eta)= \eta^2 \sum_{\ell =1}^{\infty} \frac{1}
{\ell (\ell^2+ \eta ^2)} -\log \eta - 0.57722 \cdots .
\end{equation}
In Eq.(4) $\tau_{0}$ is the phase shift due to the vacuum 
polarization potential\cite{vacpol}
\begin{equation}  
 V^{vac}(r)= \lambda \frac{e^2}{r} \int_{4 m_{e}^2}^{\infty} dt 
 \frac{ e^{-r \sqrt{t}}}{2 t} (1+ \frac{2 m_{e}^2}{t}) 
 \sqrt{1-\frac{4 m_{e}^2}{t}}
\equiv \lambda  \frac{e^2}{r} I(r) \quad ,
\end{equation}
where  $m_{e}$ is the mass of the electron and $\lambda=2 e^2/3 
\pi =1.549 \times 10^{-3}$.  Functions $\tau_{0}$, $\chi_{0}$, 
$\phi_{0}$ and $\ell_{0}(\eta)$ have the order of magnitudes of the 
vaccum polarization, and introduced in the previous paper.
 
      By using the modified effective range function $X_{0}(\nu)$, 
we define the S-wave amplitude $a_{0}(\nu)$ of the p-p scattering by
\begin{equation}  
a_{0}(\nu)=\frac{\sqrt{m^2+\nu}}{ X_{0}(\nu) - m e^2 h(\eta) -i 
\sqrt{\nu} C_{0}^2} \quad .
\end{equation}
 The relation between $a_{0}(\nu)$ and the phase shift 
 $\delta_{0}^{E}$ is obtained if we substitute $X_{0}(\nu)$ of 
 Eq.(4) into Eq.(7), and which reduces to the well-known form   
\begin{equation}  
a_{0}(\nu)=\frac{1}{C_{0}^2} \frac{\sqrt{m^2+\nu}}{\sqrt{\nu}} 
e^{i \delta_{0}^{E}(\nu)} \sin \delta_{0}^{E}(\nu) \quad ,
\end{equation}
if the functions related to the vacuum polarization are neglected.  
 The form of $a_{0}(\nu)$ of Eq.(7) is the same as that of the 
 neutron-neutron scattering $(\sqrt{m^2+\nu}/\sqrt{\nu}) e^{i \delta} 
\sin \delta$  except for the factor 
 $C_{0}^2$ given in Eq.(4), which is the penetration 
 factor.   
If we compare the S-wave amplitude of the p-p scattering of Eq.(7) 
with that of the n-n scattering Eq.(1), a combination of 
functions $(-m e^2 h(\eta) - 
i \sqrt{\nu} C_{0}^2)$ appears in place of $\ -i  \sqrt{\nu}$.  
In order to investigate the analytic structure of $a_{0}(\nu)$, 
it is convenient to rewrite the combination as
\begin{equation}   
 -m e^2 h(\eta) - i \sqrt{\nu} C_{0}^2 = -i \sqrt{\nu} + m e^2 
 \{\log (i \eta) -\psi(1+i \eta) \} \quad .
\end{equation}
Since the digamma function $\psi(z)$ has poles at non-positive 
integers, the poles on the $\eta$-plane appear on the positive 
imaginary axis. 
In terms of $\sqrt{\nu}$, which is $m e^2/(2 \eta)$, the series of 
poles appear on the negative imaginary axisis and converge to 
$\sqrt{\nu}=0$. It is the smallness of the fine structure constant 
$e^2$ and therefore of the residues of such poles that zeros of the 
denominator of Eq.(7) occur at points very close to the locations 
of the poles of $ (-m e^2 h(\eta)-i \sqrt{\nu}C_{0}^2 )$.  
Therefore the partial wave amplitude $a_{0}(\nu)$  of the p-p 
scattering has a series of poles on the second sheet of $\nu$, 
namely on the lower half plane of $\sqrt{\nu}$, whereas on the first 
sheet of $\nu$ the analytic structure of $a_{0}(\nu)$ does not change 
compared to the case of the n-n scattering.
  This fact implies that the same definition of the Kantor amplitude 
  $K_{0}(\nu)$ introduced for the neutron-neutron scattering, which 
  is given in Eq.(2), is valid also for the proton-proton scattering, 
 as long as we evaluate Im$a_{0}(\nu')$ of Eq.(2) from Eqs.(7) and 
  (9).
  Therefore the Kantor amplitude of the p-p scattering $K_{0}(\nu)$ 
  constructed in this way is free from the singularities in the 
  neighborhood of $\nu=0$, and so does not have the cut of the 
  vacuum polarization as well as that of the Coulomb interaction.        

      We can now compute the once subtracted S-wave Kantor amplitude 
minus the contribution from the one-pion exchange of the proton-proton 
scattering:
\begin{equation} 
\tilde{K}_{0}^{once}(\nu) \equiv \frac{K_{0}(\nu)}{\nu}-      
\frac{K_{0}^{1 \pi}(\nu)}{\nu} \quad .
\end{equation}

\begin{figure}[htbp]
\begin{minipage}{6.8cm}
\includegraphics[width=.99\textwidth,height=5.5cm]{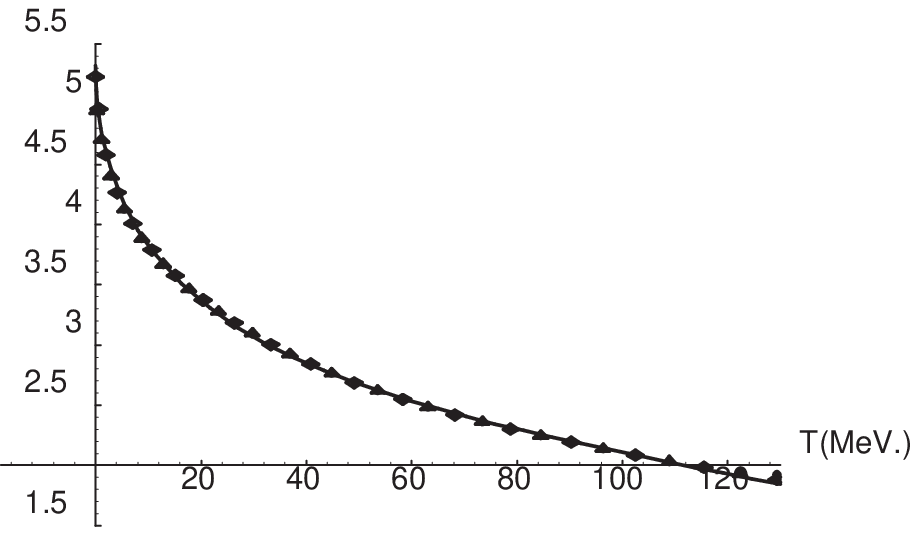}
\caption{{\small
 $-\tilde{K}_{0}^{once}(\nu)$ is plotted against $T_{lab}$ in 
  $T_{lab} < 125 \; MeV.$.    The curve is the fit by the spectrum 
  of the long range force with three parameters
  in the energy range $ 0.6\;MeV. < T_{lab} <125\; MeV.$ .  }}
\end{minipage}
\hfill
\begin{minipage}{6.8cm}
\includegraphics[width=.99\textwidth,height=5.cm]{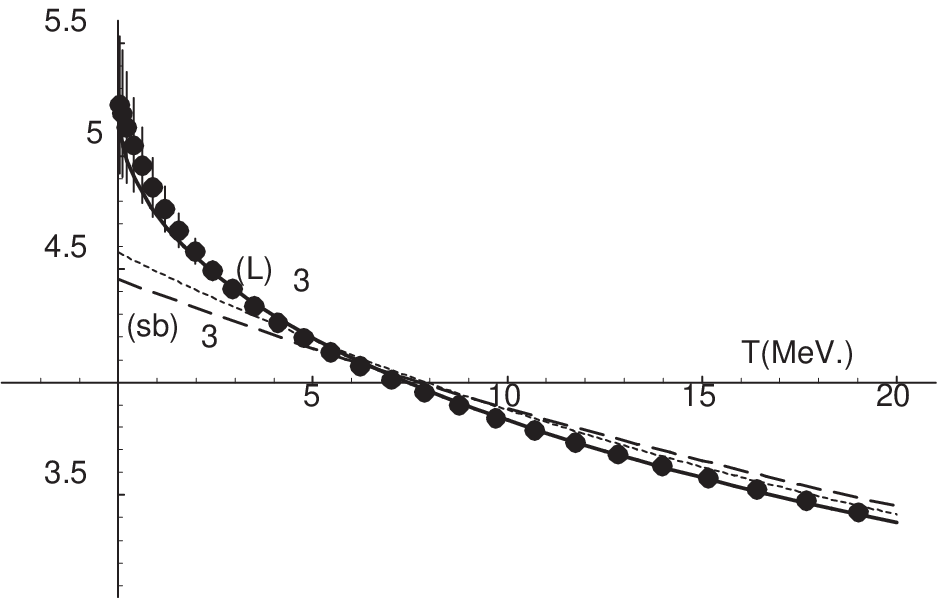}
 \caption{{\small
The enlarged graph of $-\tilde{K}_{0}^{once}(\nu)$ is plotted 
 in $ T_{lab} < 20 \; MeV.$.  The three curves are 
3-parameter fits to the data in $ 0.6\;MeV. < T_{lab} <125\;
 MeV.$.  The dotted and the dashed curves are the fits by the 
 spectra of the short range forces.
 }}
\end{minipage}
\end{figure}
In figure 1 and figure 2, $-\tilde{K}_{0}^{once}(\nu)$ is plotted 
against $T_{lab}$.  The graphs exhibit a cusp at $\nu=0$, which is 
characteristic to the attractive long range force.   The cusp is 
fitted by a spectral function of three parameters:
\begin{equation} 
A_{t}^{extra}(s,t)= \pi C' t^{\gamma} e^{-\beta t}
\end{equation} 
, and the results of the chi-square fit in $0.6 MeV. < T_{lab} <  
125 MeV.$ are
\begin{equation}  
\gamma=1.543 \quad , \qquad \beta=0.06264 \quad and \qquad 
C'=0.1762 
\end{equation}
in the unit of the neutral pion Compton wave length.
  The curve in fig.1 and $(L)_{3}$ curve in fig.2 are the fits by the 
 spectral function of the long range force $A_{t}^{extra}(s,t)$ given 
in Eqs.(11) and (12), and the $\chi$-value per data point is 0.441.  
On the other hand, other curves in fig.2  are the 
three parameter fits by the spectral function of the short range 
forces:
\begin{equation} 
 A_{t}(s,t)= \sum_{i=1}^{3} c_{i} \delta (t-t_{i}) \quad ,
\end{equation} 
where $c_{i}$ are free parameters and three $t_{i}$ are 4, 9 and 16 
for the curve $(sa)_{3}$ (dotted curve), whereas $t_{i}$ are 9, 16 
and 25 for the curve $(sb)_{3}$ (dashed curve), and the $\chi$-values 
of the fits per data points are 1.82 and 3.11 respectively.  
The curves 
indicate that the short range spectra, which mimic the spectrum of 
the two-pion exchange, cannot reproduce the cusp of  
$-\tilde{K}_{0}^{once}(\nu)$ shown in figures 1 and 2.   Details of 
the fits are found in the previous paper.\cite{prev}

\section{Interference pattern of the cross section of p-p 
scattering}
\ \ \ \ \  In the previous section, we observed a cusp at $\nu=0$ in 
$-\tilde{K}_{0}^{once}(\nu)$ which arises from the partial wave 
projection of the singular term $C'(2 \nu)^{\gamma} (1-z)^{\gamma}$. 
However the same singularity can also be observed in the angular 
distribution for fixed $\nu$.   The aim of this paper is to propose 
to observe the singularity at $z=\pm 1$ and to confirm the existence 
of the strong Van der Waals force in the p-p scattering.   Since the 
Coulomb potential also gives rise to the poles at $z=\pm 1$, we 
can expect to observe the characteristic interference pattern of 
the singular behaviors in the cross section.   In this section, 
we shall compute such a pattern by using the parameters of the 
long range force determined in the previous section. 
If we consider that the energy dependent phase shift data are 
extracted from the measurements of different laboratories, the 
observation of the singular behavior in the angular distribution 
is the more direct way to confirm the existence of the long 
range force.

    The amplitude due to the extra spectrum of Eq.(11) is 
\begin{eqnarray} 
& &\frac{m}{\sqrt{\nu}}  W(\nu,z)  =  \frac{\pi C'}{\pi} 
\int_{0}^{\infty} dt'
  \frac{t'^{\gamma} e^{-\beta t'}}{t'+2 (1-z)} \nonumber \\
       & = & C' (2 \nu)^{\gamma} (1-z)^{\gamma} \Gamma (\gamma +1)   
\Gamma(-\gamma, 2 \nu \beta (1-z) , \infty) \exp [ 2 \nu \beta (1-z)]
\end{eqnarray}
, where $\Gamma (x, a, b)$ is the incomplete gamma function defined 
by $\int_{a}^{b} t^{x-1} e^{-t} dt$.
The cross section $\sigma (\theta)$ of the proton-proton scattering in 
the low energy region is\cite{kiang}
\begin{eqnarray} 
\nu \sigma( \theta)&=&\frac{1}{4} |\tilde{f}^{s} (\theta) + 2 \exp[i 
\delta_{0}^{E}] \sin \delta_{0}^{E} + \alpha_{e}(z)|^2 + 
\nonumber \\  
&+& \frac{3}{4} |\tilde{f}^{t} (\theta) +6 \delta_{1,C} z 
 + \alpha_{o}(z)|^2 + \Delta  \quad .
\end{eqnarray}
$\tilde{f}^{s} (\theta)$ and $\tilde{f}^{t} (\theta)$ are the 
Coulomb amplitude plus the one-pion exchange contribution of 
the spin-singlet and the spin-triplet states respectively, and 
they are 
\begin{eqnarray} 
\tilde{f}^{s} (\theta)&=&-\frac{\eta}{2} \{ \frac{1}{S^2}\exp[-i \eta
 \log S^2]+\frac{1}{C^2} \exp[-i \eta \log C^2] \} + \nonumber \\
 &+& f^2 m \sqrt{\nu} \{(\frac{1}{1-t}+\frac{1}{1-u})-\frac{2}
 {2 \nu} Q_{0}(1+\frac{1}{2 \nu}) \}
\end{eqnarray}
and
\begin{eqnarray} 
\tilde{f}^{t} (\theta)&=&-\frac{\eta}{2} \{ \frac{1}{S^2}\exp[-i \eta
 \log S^2]-\frac{1}{C^2} \exp[-i \eta \log C^2] \} - \nonumber \\
&-& \frac{f^2}{3} m \sqrt{\nu}\{(\frac{1}{1-t}+\frac{1}{1-u})-\frac{6}
 {2 \nu} Q_{1}(1+\frac{1}{2 \nu}) z \} \quad ,
\end{eqnarray}
, in which 
\begin{equation} 
S=\sin \frac{\theta}{2} \quad , \qquad C=\cos \frac{\theta}{2} 
\qquad and \qquad \eta=\frac{e^2}{\hbar c \beta_{L}} \quad ,
\end{equation}
where $\beta_{L}$ is the velocity in the laboratory system and it is 
written in terms of $\nu$ 
\begin{equation} 
\beta_{L}=\frac{2 \sqrt{\nu} \sqrt{1+\frac{\nu}{m^2}}}
{m (1+\frac{2 \nu}{m^2})} \quad .
\end{equation}

  In Eq.(15) $\alpha_{e}(z)$ and $\alpha_{e}(z)$ are the even and odd 
functions of $z$ which take care of the contributions from the 
higher partial waves and for very small $\nu$ they are negligible.
When the hadron interaction is short range, 
since the poles of the one-pion exchanges are already separated, 
the nearest singulary of $\alpha (z)$'s on the $z$-plane occurs at
\begin{equation} 
 z= \pm (1 + \frac{t_{min}}{2 \nu} )
 \end{equation}
 with $t_{min}=4$, the threshold of the spectrum of the two-pion 
 exchange.
 As an example, we shall fix the incident energy at $T_{lab}=$20 MeV.,
  namely at $\nu=0.515$.  In such a case, the analytic domain on the 
$z$-plane is the Lehmann ellipse whose semi-major axis is around 5 
and foci are at $z= \pm 1$.    However if the two-pion exchange 
spectrum is very small in the threshold region and the main 
cotribution comes from the 
$\sigma$-meson (560 MeV.), then $t_{min}=16$ and the semi-major axis 
is around 17.  For $\alpha_{e}(z)$ it is sufficient to retain only 
the S and D waves to reproduce the amplitude within the error 
$10^{-3}$ in the physical region $-1 \leq z \leq 1$.    
For the odd function $\alpha_{o}(z)$ we shall consider 
two cases, the first one is to retain only the P wave, whereas the 
second one is to retain the P and F waves.     
 
     To observe the extra singularity at $z= \pm 1$, we firstly 
determine the coefficients of the polynomials $\alpha_{e}(z)$ and 
$\alpha_{o}(z)$ using the cross section in the off-forward region 
$ -z_{1} \leq z \leq z_{1}$.  Notation $\sigma_{smooth}(z)$ will 
be used 
for the cross section which is obtained by using the polynomials 
thus determined, because this is the smooth continuation of the 
cross section from the off-forward region to whole the physical 
region $ -1 \leq z \leq 1$.  When the nuclear forces are short 
range, the observed cross section must coincide with 
$\sigma_{smooth}(z)$ in the forward region $ z_{1} \leq z \leq 1$ 
within the error.
On the other hand, if the long range force exists and the extra 
singularity is of the form $(1-z)^{\gamma}$, then the observed 
cross section deviates from $\sigma_{smooth}(z)$ in the forward 
region.   We can evaluate such a deviation if the spectral 
functions of the long range force Eqs.(11) and (12) are given. 

     By using the amplitude of the long range force $W(\nu,z)$ 
introduced in Eq.(14), $\alpha(z)$'s are written as
\begin{equation} 
\alpha_{e}(z)=(W(\nu,z)+W(\nu,-z)-2 w_{0}-2 w_{2} z^2)
+c'_{0}+c'_{2} z^2
\end{equation}
and
\begin{equation} 
\alpha_{o}(z)=(W(\nu,z)-W(\nu,-z)-2w_{1} z)+c'_{1} z \quad ,
\end{equation}
 in which $w_{i}$'s are chosen in such a way that $(w_{0}+w_{1} z+
 w_{2} z^2)$ becomes the best fit to $W(\nu,z)$ in the given 
 off-forward region $ -z_{1} \leq z \leq z_{1}$.  In particular 
 for $z_{1}=0$, $w_{0}$, $w_{1}$ and $w_{2}$ are the value, slope 
 and curvature at $z=0$ respectively.   With these $\alpha(z)$'s 
we can compute the cross section $\sigma_{long}(z)$ from Eq.(15). 
 The relative deviation of the cross section $\Delta \sigma/\sigma$
is defined by
\begin{equation} 
\Delta \sigma/\sigma (\theta)=\frac{\sigma_{long}(z)-
\sigma_{smooth}(z)}{\sigma_{smooth}(z)}
\end{equation}
and if we retain only the first order of the small quantities 
$\alpha_{e}(z)$, $\alpha_{o}(z)$ and $\Delta$ in Eq.(15), 
the relative deviation reduces to 
\begin{eqnarray} 
\Delta \sigma/\sigma (\theta)&=&\tilde{g}^{s} (\theta)  
(W(\nu,z)+W(\nu,-z)-2 w_{0}-2 w_{2} z^2)+  \nonumber \\
&+& \tilde{g}^{t} (\theta)  
(W(\nu,z)-W(\nu,-z)-2 w_{1} z)  \quad , 
\end{eqnarray}
where
\begin{equation} 
\tilde{g}^{s} (\theta)=\frac{1}{2 \sigma_{0}(\theta)} {\rm Re}
(\tilde{f}^{s} (\theta) + 2 \exp[i 
\delta_{0}^{E}] \sin \delta_{0}^{E}) 
\end{equation}
and
\begin{equation} 
\tilde{g}^{t} (\theta)=\frac{3}{2 \sigma_{0}(\theta)} {\rm Re}
(\tilde{f}^{t} (\theta) +6 \delta_{1c} z)  \quad .
\end{equation}
In Eqs.(25) and (26), $\sigma_{0}(\theta)$ is the zeroth order 
of the 
cross section, and obtained by setting the samall quantities 
$\alpha_{e}(z)$, $\alpha_{o}(z)$ and $\Delta$ in Eq.(15) equal to 
zero.

      In figure 3 and 4, the coefficient functions $\tilde{g}^{s} 
 (\theta)$ and $\tilde{g}^{t} (\theta)$ are displayed, in which 
 we use the phase shifts $\delta_{0}^{E}=50.96^{\circ}$ and 
 $\delta_{1c}=0.516^{\circ}$ at $T_{lab}=$ 20 MeV. , and also the 
 $\pi$-N coupling constant $g^2/4 \pi=14.4$ where $g^2/4 \pi=4 
 m^2 f^2$.

\begin{figure}[htbp]
\begin{minipage}{6.8cm}
\includegraphics[width=.99\textwidth,height=5.5cm]{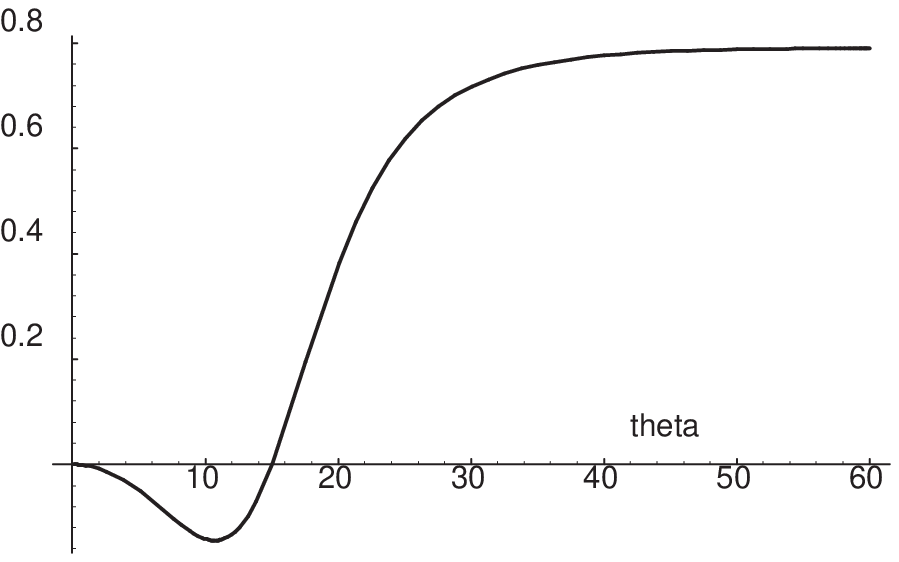}
\caption{{\small
  The singlet coefficient function $\tilde{g}^{s} 
 (\theta)$ is plotted against $\theta$.  $T_{lab}=20$ MeV. }}
\end{minipage}
\hfill
\begin{minipage}{6.8cm}
\includegraphics[width=.99\textwidth,height=5.cm]{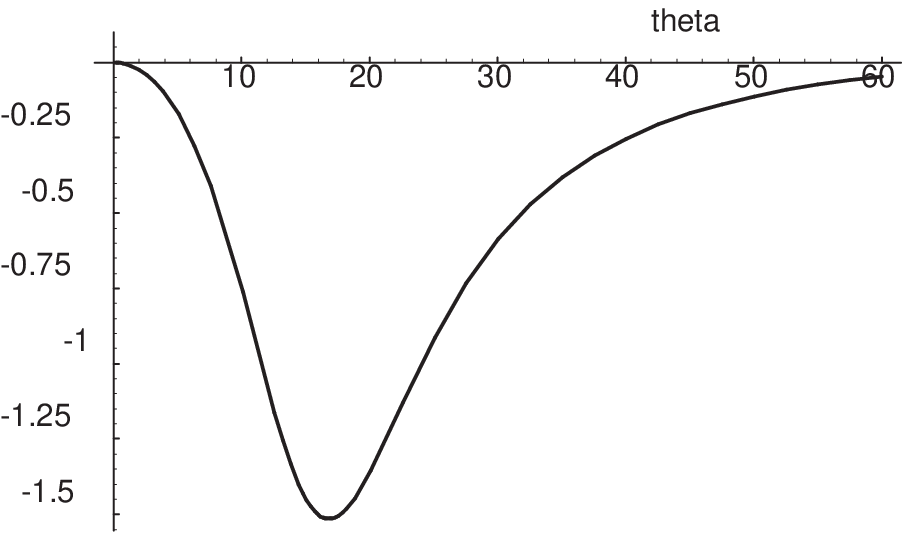}
 \caption{{\small
The triplet coefficient function $\tilde{g}^{t} 
 (\theta)$ is plotted against $\theta$.  $T_{lab}=20$ MeV.
 }}
\end{minipage}
\end{figure}

In figure 5, the relative deviation $\Delta \sigma/\sigma (\theta)$ 
is plotted against $\theta$, the scattering angle in the center of 
mass system.   The numbers attached to the curves are $\theta_{1}$ 
the boundary of the off-forward region $ \theta_{1} \leq \theta 
\leq 180^{\circ}-\theta_{1}$, in which the parameters in $\alpha(z)$'s 
are determined.   The curves show narrow dips at $\theta=15^{\circ}$ 
with the depth around 1 per cent.   This is our prediction of the 
interference pattern derived from the parameters of the long range 
force.  If the long range force does not exist, the curves of the 
relative deviation $\Delta \sigma/\sigma (\theta)$ must be zero 
within the error.

\begin{figure}[htbp]
 \includegraphics[width=.9\textwidth,height=7.0cm]{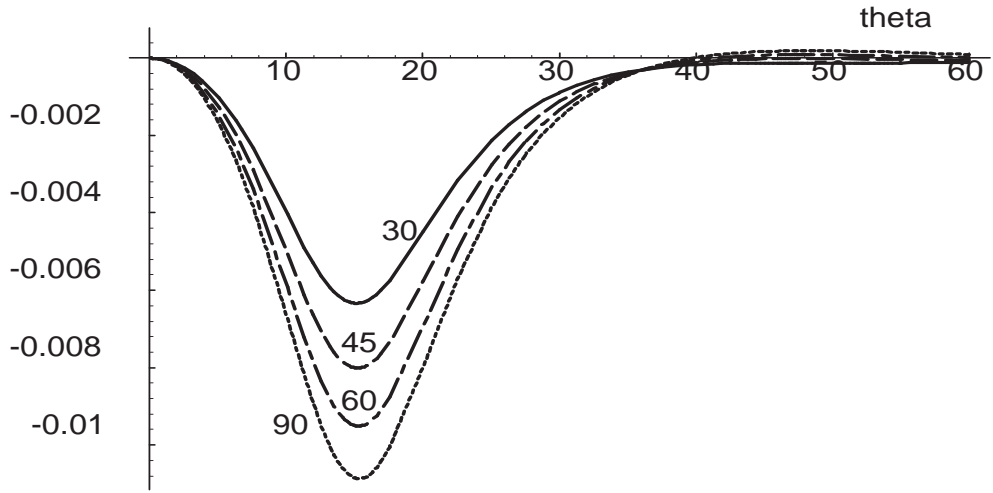}
\caption{{\footnotesize
The relative deviation $\Delta \sigma/\sigma (\theta)$ is plotted 
against $\theta$ for $T_{lab}=20$ MeV..  Three parameters are 
determined in the off-forward region $\theta_{1} \leq \theta 
\leq 180^{\circ}-\theta_{1}$.  The curves correspond to 
$\theta_{1}=30^{\circ}$, $45^{\circ}$, $60^{\circ}$ and 
$90^{\circ}$ respectively.  }}
\end{figure}

In figure 6, the curves $\Delta \sigma/\sigma (\theta)$, in which 
the F-wave as well as the P-wave term in $\alpha_{o}(z)$ are 
retained, are shown.  The dips appear at $\theta=13^{\circ}$.

\begin{figure}[htbp]
 \includegraphics[width=.9\textwidth,height=7.0cm]{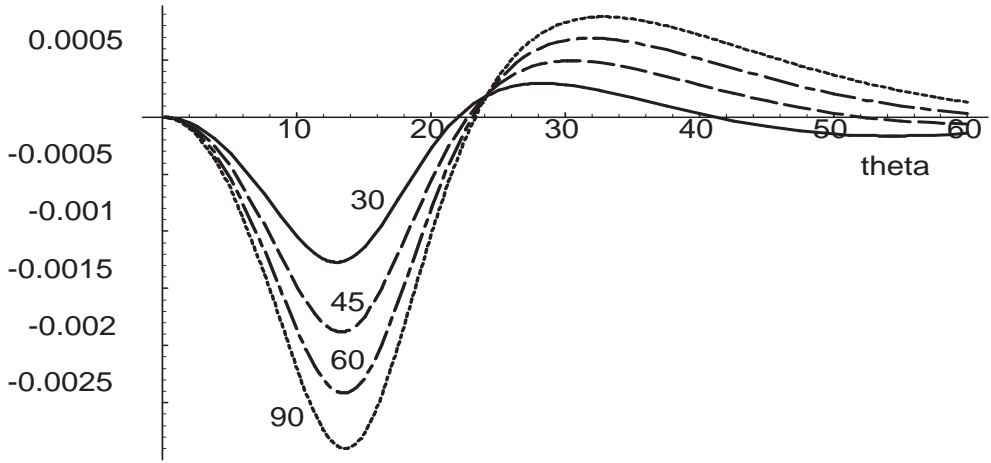}
\caption{{\small
The relative deviation $\Delta \sigma/\sigma (\theta)$ is plotted 
against $\theta$ for $T_{lab}=20$ MeV..  Four parameters are 
determined in the off-forward region $\theta_{1} \leq \theta 
\leq 180^{\circ}-\theta_{1}$.  The curves correspond to 
$\theta_{1}=30^{\circ}$, $45^{\circ}$, $60^{\circ}$ and 
$90^{\circ}$ respectively.  }}
\end{figure}
 
 \section{Remarks and Comments}
\ \ \ \ \ In this paper we propose to measure precisely 
the angular distribution 
of the cross section of the low energy proton-proton scattering, 
in order to confirm the long range interaction 
in the nuclear force.  By using the parameters of the spectrum of 
the long range force obtained from the analysis of the S-wave 
phase shift of the p-p scattering, we predict the characteristic 
interference pattern in the angular distribution of the p-p cross 
section, which has a dip at $\theta_{c.m.}=14^{\circ} $ with the 
depth around 1 \%.     The observation of such a pattern is the 
more direct way to confirm the long range force, because the energy 
dependent phase shifts are obtained from the data of different 
laboratories by constructing a consistent curve, in which the 
correction factor of the incident beam is assigned  
to each experiment.   If we consider the precision of the measurement 
the low energy proton-proton is an ideal place to observe the strong 
Van der Waals force.

     Another good place to observe the Van der Waals force is the 
low energy ($T_{lab} \sim$ 1 MeV.) neutron-Pb scattering.\cite{npb} 
 This is 
because the strength of the long range potential is magnified by a 
factor A, the mass number, and it is relatively easy to observe the 
anomaly of the angular distribution.   Since the Van der Waals 
force is universal, we can expect to observe such a force also in 
other processes such as in the $\pi$-$\pi$ scattering.\cite{universal} 
 Although the 
precisions of the $\pi$-$\pi$ data are not very high, the P-wave 
amplithde of the $\pi$-$\pi$ is another place easy to observe the 
strong Van der Waals force, because we can compute the spectral 
function of the two-pion exchange from the $\pi$-$\pi$ data, and by 
removing the spectrum from the amplitude we can prepare 
the wide domain of analyticity.  If all the forces are short 
range, the Kantor amplitude must be almost constant.\cite{kpipi}  

\begin{figure}[htbp]
 \includegraphics[width=.9\textwidth,height=7.0cm]{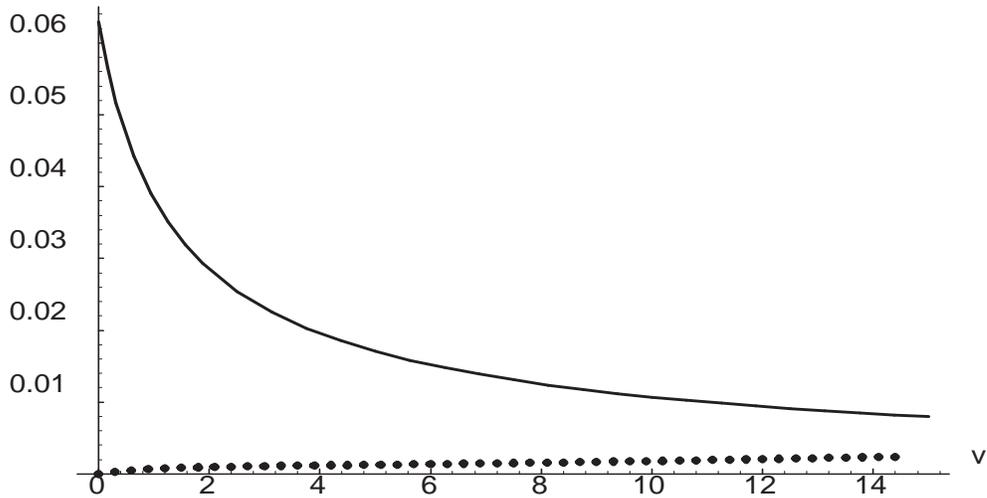}
\caption{{\footnotesize
 The once subtracted Kantor amplitude of the P-wave of the $\pi$-
 $\pi$ scattering $K_{1}(\nu)/\nu$ is plotted against $\nu$.
 The dotted curve is $-K_{1}^{2 \pi}(\nu)/\nu$, the two-pion 
 exchange contribution.  If all the forces are short range, 
 two curves must coincide up to a constant.   }}
\end{figure}

In figure 7, 
the once subtracted Kantor amplitude $K_{1}(\nu)/\nu$ of the P-wave 
pi-pi scattering is shown along with the contribution from the 
two-pion exchange spectrum.   Here again the cusp of the attractive 
sign appears.   By making the chi-square fit, the parameters are 
determined.
\begin{equation} 
 \gamma=1.95 \quad , \qquad C'=0.0161  \quad and \qquad 
 \beta=0.144
 \end{equation} 
The long range force in $\pi$-$\pi$ is close to the Van der Waals 
force of the Casimir-Polder type rather than the London type.

 \end{document}